\documentclass[amssymb, amsmath, groupedaddress, reprint]{revtex4-1}

\usepackage{graphicx}
\usepackage[colorlinks=true,linkcolor=blue]{hyperref}
\usepackage{soul,xcolor}
\setstcolor{red}

\begin{document}

\title{Active Underwater Detection with an Array of Atomic Magnetometers}

\author{Cameron Deans}
\author{Luca Marmugi}
\email{Corresponding author: l.marmugi@ucl.ac.uk}
\author{Ferruccio Renzoni}

\affiliation{Department of Physics and Astronomy, University College London, Gower Street, London WC1E 6BT, United Kingdom}

\date{}


\begin{abstract}
We report on a 2$\times$2 array of radio-frequency atomic magnetometers in magnetic induction tomography configuration. Active detection, localization, and real-time tracking of conductive, non-magnetic targets are demonstrated in air and saline water. Penetration in different media and detection are achieved thanks to the sensitivity and tunability of the sensors, and to the active nature of magnetic induction probing. We obtained a 100$\%$ success rate for automatic detection and 93$\%$ success rate for automatic localization in air and water, up to 190~mm away from the sensors' plane (100~mm underwater). We anticipate magnetic induction tomography with arrays of atomic magnetometers finding applications in civil engineering and maintenance, oil\&gas industry, geological surveys, marine science, archeology, search and rescue, and security and surveillance.
\end{abstract}

\maketitle

\onecolumngrid
\vskip 5pt
\begin{center}
\textit{This is a preprint version of the article appeared in Appl. Opt. {\bf 57}, 10, 2346-2351 (2018) DOI: \href{https://doi.org/10.1364/AO.57.002346}{10.1364/AO.57.002346}.}
\end{center}

\vskip 10pt

\twocolumngrid
\section{Introduction}
Detection, localization, and tracking of remote or concealed objects is an open problem in many fields, in particular when penetration through absorbing or scattering media is required \cite{comm, geo, archeo, rescue, securitytracking, uav}. A critical case is underwater detection, where many conventional detection techniques are often ineffective \cite{nist2017}. For example, optical methods can be severely challenged \cite{schettini2010}, and acoustic systems incur increased difficulties at shallow depths \cite{shallowwaterchallenge, shallowwaterchallenge2}. In addition, the use of ionizing radiation or particles is often impractical, or technically impossible because of the interaction (or the lack thereof) with media and targets.

Here, we present a proof-of-concept demonstration of detection and localization with an array of atomic magnetometers (AMs) \cite{budker} operating in magnetic induction tomography (MIT) \cite{griffithsmit} configuration. This new approach does not suffer from the abovementioned limitations, and it is suitable for multi-purpose use. Thanks to the sensitivity and tunability of AMs, MIT can be implemented with low fields and at low frequencies, thus matching the requirements for long range penetration in media. \textcolor{black}{In addition, the room temperature operation of AMs in unshielded environments, low running costs, miniaturization, scalability, and low costs for additional units make them an ideal solution for remote or underwater detection and localization \cite{nist2017}}.

We demonstrate active and automatic detection, localization, and tracking of conductive non-magnetic targets in air and in saline water, by continuously and simultaneously monitoring the output of a 2$\times$2 planar array of AMs. We have obtained an overall success rate for detection of 100$\%$ in the explored configurations, and 93$\%$ successful localization. Real-time tracking of moving targets is also demonstrated, as well as multiple target simultaneous detection.

Although arrays of AMs have been previously realized (see, for example, \cite{array2009,array2012,array2016}), their operation in MIT modality has not. Our results demonstrate the feasibility of MIT with an array of AMs, and the relevance of such an instrument for detection and localization of concealed targets. The technology can \textcolor{black}{also be} integrated with remote detection of rotating machinery with AMs \cite{ao}, to create a multi-function sensing platform. This would provide a compact, safe, and active alternative for remote monitoring, surveying, and surveillance in many fields, as well as for increasing the throughput of MIT-AM imaging systems \cite{apl,opex}.

\section{Array of Radio-frequency Atomic Magnetometers}
\label{sec:array}

\begin{figure*}[htbp]
\centering
\includegraphics[width=\linewidth]{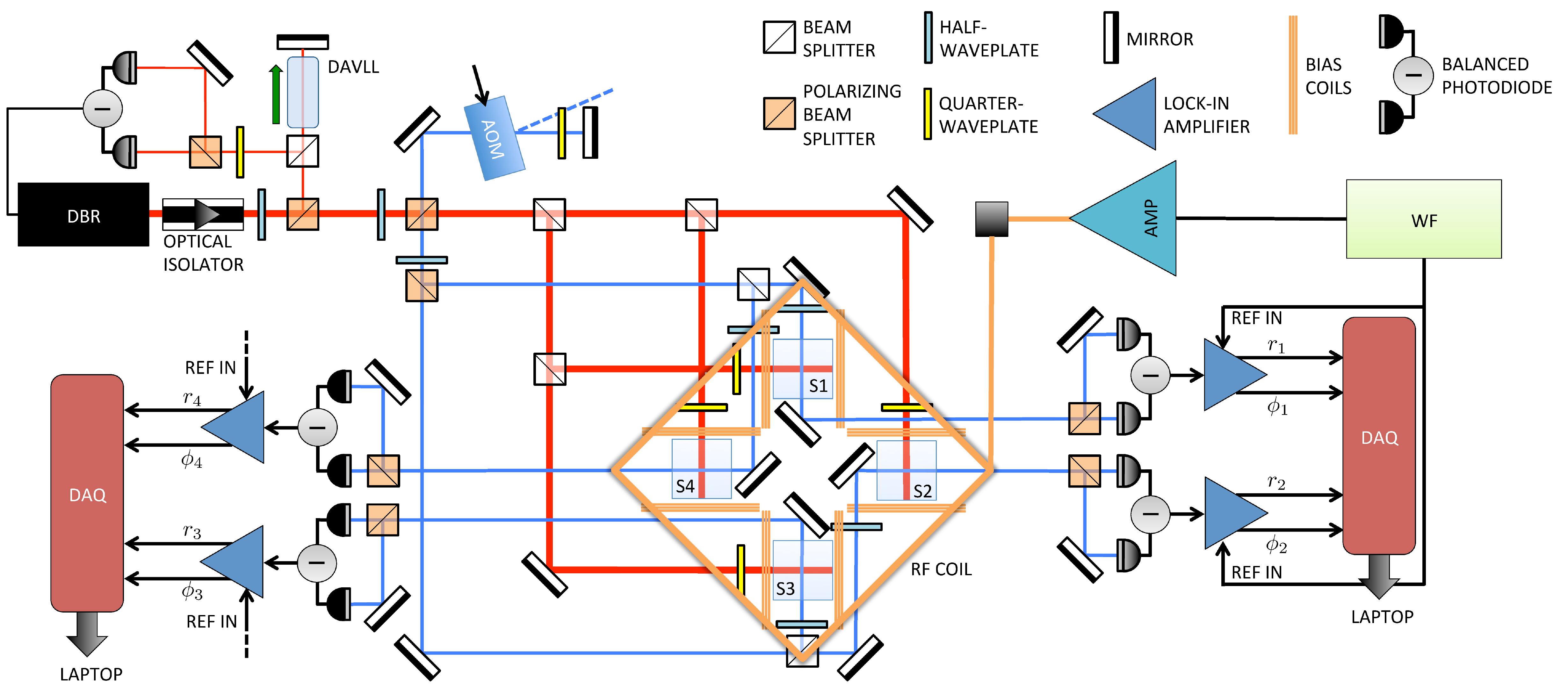}
\caption{Simplified sketch of the 2$\times$2 RF AMs array for detection and localization. DBR: distributed Bragg reflector laser. DAVLL: dichroic atomic vapor laser lock. AOM: acousto-optic modulator. DAQ: data acquisition board. Amp: current amplifier. WF: waveform generator. REF IN: reference input. $r_{n}$ and $\phi_{n}$ are the amplitude and phase signals, respectively, of the n-th sensor Sn.}
\label{fig:setup}
\end{figure*}

The setup is sketched in Fig.~\ref{fig:setup}. Four radio-frequency (RF) atomic magnetometers  (AM) \cite{rfam1, rfam2, witold}, labelled as Sn, where n=1,...,4, are arranged in a 2$\times$2 planar configuration. Each sensor relies on Faraday rotation to detect the presence of a conductive target, where eddy currents are induced by an AC magnetic field orthogonal to the sensors' plane. The output of each sensor (2 channels) is \textcolor{black}{multiplexed in a DAQ board (NI USB-6009)} and analyzed in real time \textcolor{black}{via} LabVIEW.  Automatic control and variable thresholds (i.e. ``decision levels'') are included for alarm triggering.

Detection is based on MIT. An AC primary magnetic field induces eddy currents in the target, which in turn generate a secondary field, oscillating at the same frequency. A phase-sensitive detection scheme referenced to the primary field extracts the amplitude of the secondary field and its phase lag $\phi$. The presence of a target is detected by measuring perturbations to the total magnetic field. This approach is inherently active: it triggers an unavoidable response in the target. The technique does not rely on intrinsic magnetic signatures, and is therefore well-suited for non-magnetic or de-magnetized targets. Furthermore, the electromagnetic near-field nature of the process allows operation also in shallow water, where acoustic-based systems are challenged.

\begin{figure}[htbp]
\centering
\includegraphics[width=\linewidth]{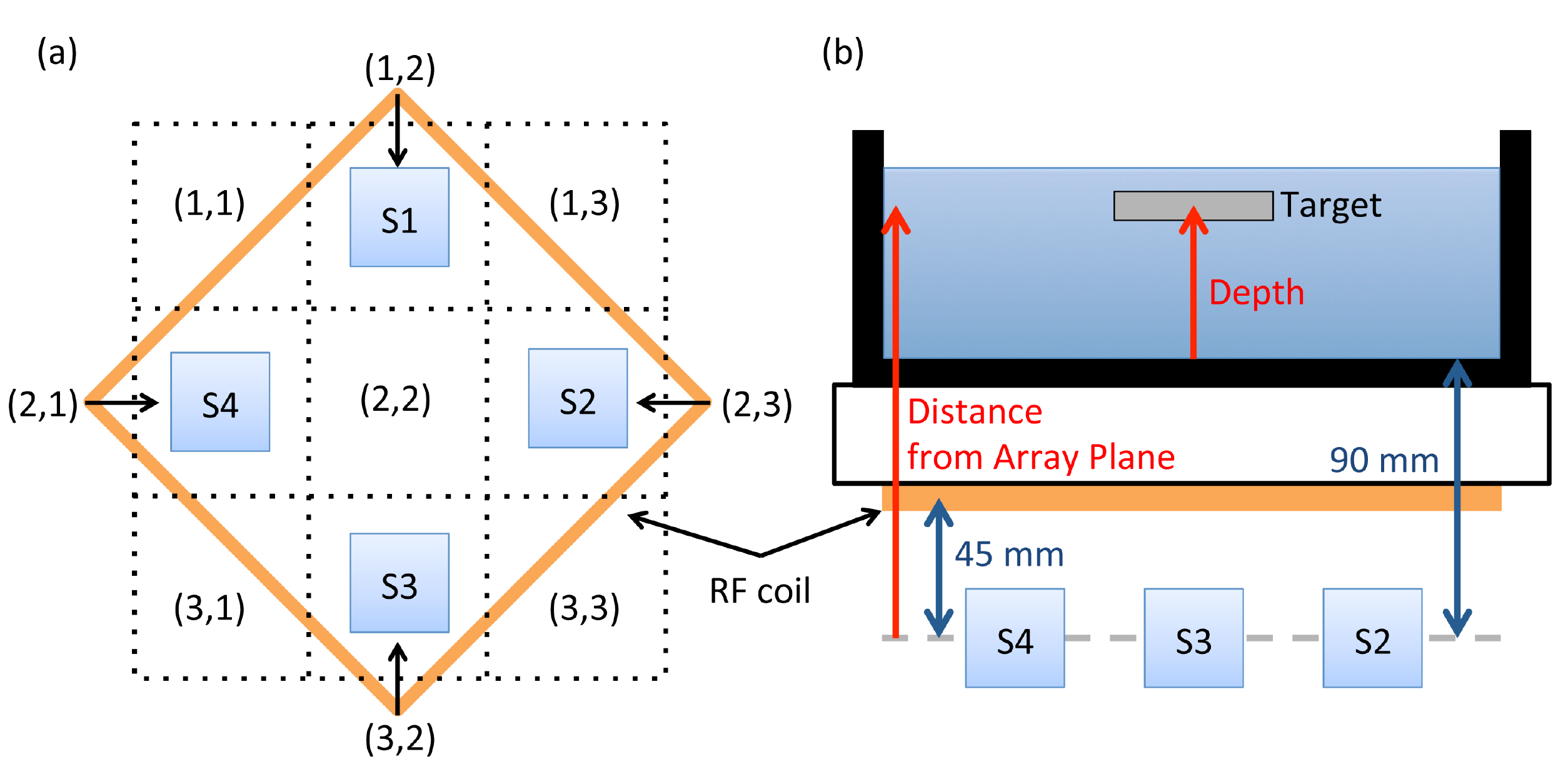}
\caption{Arrangement for detection and localization. \textbf{(a)} Coordinate grid, parallel to the sensors' plane. Each square is 77.5$\times$77.5~mm$^{2}$. \textbf{(b)} Arrangement of sensors, RF coil, target, and - when applicable - \textcolor{black}{saline} water.}
\label{fig:arrangement}
\end{figure}

In our setup, the primary field is produced by a square coil (RF coil) of side L=230~mm (15 turns), placed in a parallel plane 45~mm above the sensors' plane (Fig.~\ref{fig:arrangement}). The coil is powered by a bipolar amplifier driven by a waveform generator. 

The four RF AMs are based on a crossed pump/probe configuration, exciting the D$_{2}$ line of $^{87}$Rb. The atomic vapor is contained in 25~mm cubic quartz cells, with 20~Torr of N$_{2}$ as buffer gas. A single semiconductor laser source at 780~nm generates four $\sigma^{+}$ polarized probe beams (1~mW, beam waist 4~mm, 4.6I$_{sat}$) tuned to the F=2 $\rightarrow$ F$^{\prime}$=3 transition. Circular Helmholtz coils (diameter 85~mm) provide a bias field for optical pumping for each Sn.

The MIT primary field acts as an RF drive for orthogonal components of the atomic polarization, by resonantly coupling adjacent Zeeman sub-levels. This sets the atomic spins in precession (Larmor precession) at the frequency of the primary field. Such motion is detected through polarization rotation by four $\pi$-polarized probe beams (50~$\mu$W, beam waist 2.5~mm), detuned by +420~MHz with respect to the ``pump'' transition. Faraday rotation is measured by four independent polarimeters, whose output is selectively amplified by four dual-phase lock-in amplifiers. This allows simultaneous operation of the four sensors. 

The secondary field produced by the target causes a change in the polarization rotation and therefore a change in the amplitude r$_{n}$ and in the phase $\phi_{n}$ of the n-th sensor's output. In this way, the 2$\times$2 array provides 8 streams of data for analysis.

The four sensors are arranged at the \textcolor{black}{vertexes} of a square with side 105~mm. This distance and the mutual orientation allowed reduction of cross-talk effects to a negligible value, as well as a satisfactory coverage of the experimental area. Each sensor has a sensitivity of 3~nT/$\sqrt{\mbox{Hz}}$ at 10~kHz, measured with \textcolor{black}{all four sensors} in simultaneous operation. The sensors' performance could be further improved by active compensation of stray magnetic fields \cite{opex}. Unlike previous realizations of MIT with RF AM single sensors \cite{apl, wickenbrock, opex}, the targets are not enclosed between the sensor(s) and the RF coil.  The \emph{entire} sensing system lies below (or above) the object of interest. This has major advantages in view of practical applications, from screening \cite{opex} to biomedical imaging \cite{scirep}, as well as underground and underwater surveying.

\section{Array Operation: Detection and Localization of Conductive Targets}
The penetration of an AC magnetic field in media is governed by the exponential decay $B(z)=B(z=0)\exp(-z/\delta(\nu))$, where the skin depth $\delta(\nu)$ is: 

\begin{equation}
\delta(\nu) =\dfrac{1}{2\pi \nu\sqrt{\dfrac{\mu\varepsilon}{2} \left( \sqrt{1+\left( \dfrac{\sigma}{2\pi\varepsilon\nu}\right)^{2}}-1\right)}}~.\label{eqn:delta}
\end{equation}

\noindent Here, $\nu$ is the AC field's frequency; $\varepsilon$ is the permittivity; $\mu$ is the permeability; and $\sigma$ is the electric conductivity of the medium. According to Eq.~\ref{eqn:delta}, low frequencies are required for long range penetration and hence remote detection. In this regime, AMs outperform conventional sensors \cite{savukov, nist2017}. A primary field frequency of between 10~kHz and 20~kHz is chosen for this work. This allows sufficient penetration of the MIT fields in media and targets, with negligible attenuation in air ($\sigma_{air}\approx 10^{-15}$~S/m \cite{air}): \textcolor{black}{penetration exceeding 10$^{8}$ km can be obtained in principle. This range is significantly reduced in sea water, as discussed in Sec.~4.\ref{subsec:underwater}}. We recall that MIT operation of RF AM in a band as low as 10$^{2}$~Hz has been recently reported \cite{opex}. \textcolor{black}{With this choice of frequencies and distances between the sensor array and the target, the electromagnetic interaction is limited to the near-field regime. This corresponds to the quasi-static limit of electromagnetism, described by diffusion equations, rather than to the familiar wave propagation regime of far-field electromagnetism}.

\begin{figure}[ht]
\centering
\includegraphics[width=\linewidth]{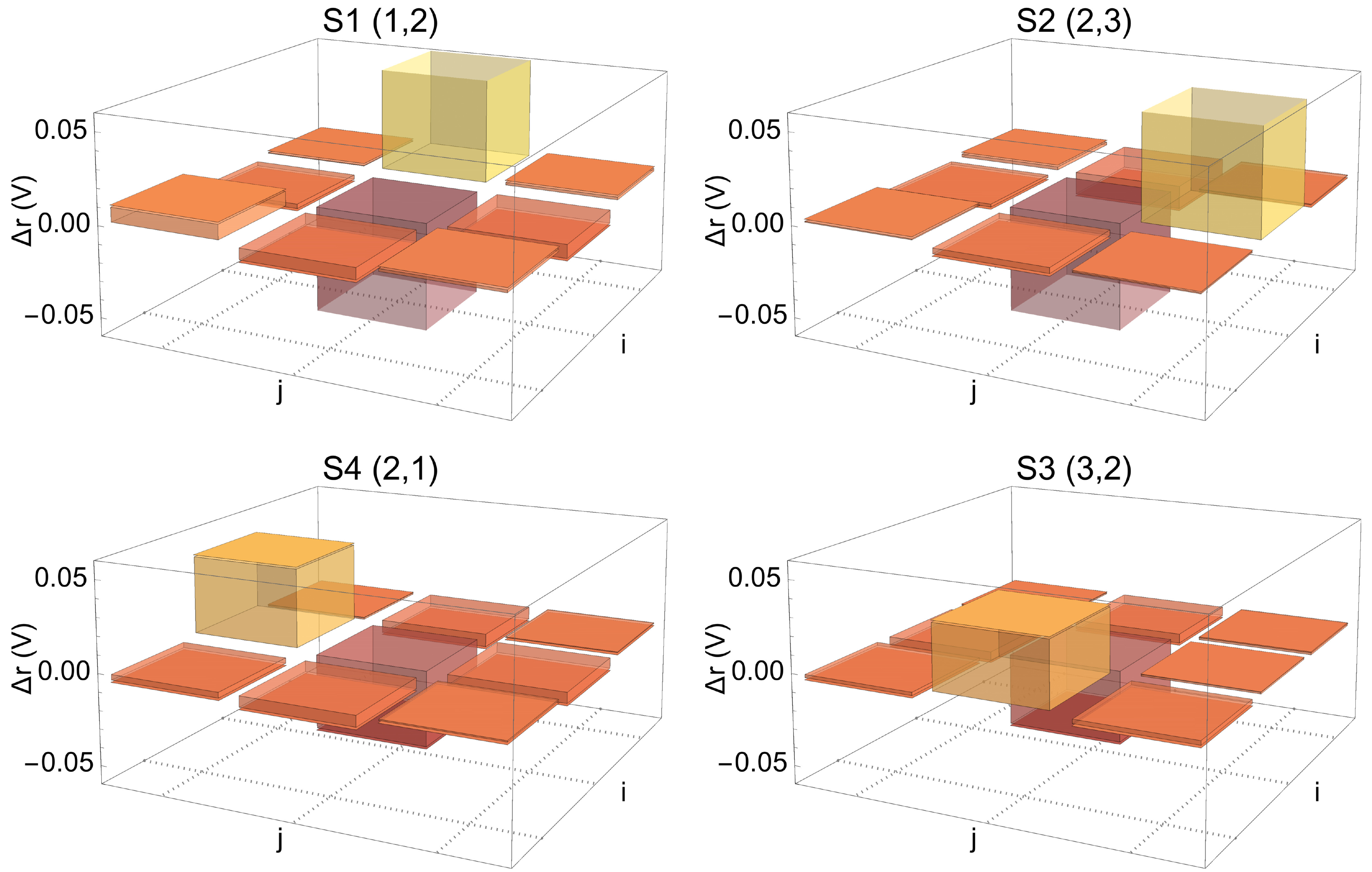}
\caption{Response of the $n$-th sensor (Sn ($i_{n}, j_{n}$)):  $\Delta$r produced by an Al plate (105$\times$110$\times$10~mm$^{3}$) in air 90~mm above the sensors' plane, detected at 20~kHz. The plate is placed in each of the 9 grid positions and the corresponding $\Delta$r is independently recorded with each sensor. \textcolor{black}{Different responses may be observed due to the independent optimization of each sensor in the array.}}
\label{fig:response-r}
\end{figure}

To demonstrate the operation of the array and the limited impact of cross-talk, the response of the four sensors is measured as a function of the target's position. Figure~\ref{fig:response-r} shows the variation in the amplitude of the four Sn when an Al plate (105$\times$110$\times$10~mm$^{3}$) is placed on the nine positions $(i,j)$. $\Delta$r=|r$_{target}$-r$_{bg}$|  (where r$_{target,bg}$ indicate the amplitude of the Faraday rotation signal with and without the target, respectively) is plotted as a function of target's position.

A systematic increase of $\Delta$r when the target is in proximity of the sensor is observed. Given the experimental arrangement (Fig.~\ref{fig:arrangement}(b)), direct screening of the primary field by the target is excluded. This behavior is consistent with MIT detection: the secondary field excited in the target perturbs the RF-driven Faraday rotation in the array. The effect is larger when the target is closer to the sensor, where the secondary field is stronger: up to 10 times larger than the neighboring values. Furthermore, a systematic decrease of the signal from all sensors is observed when the object is in the center of the grid (2,2). This allows unambiguous localization of the target in five different positions. \textcolor{black}{The four vertexes of the coordinate grid are not taken into consideration.}

\begin{figure}[htbp]
\centering
\includegraphics[width=\linewidth]{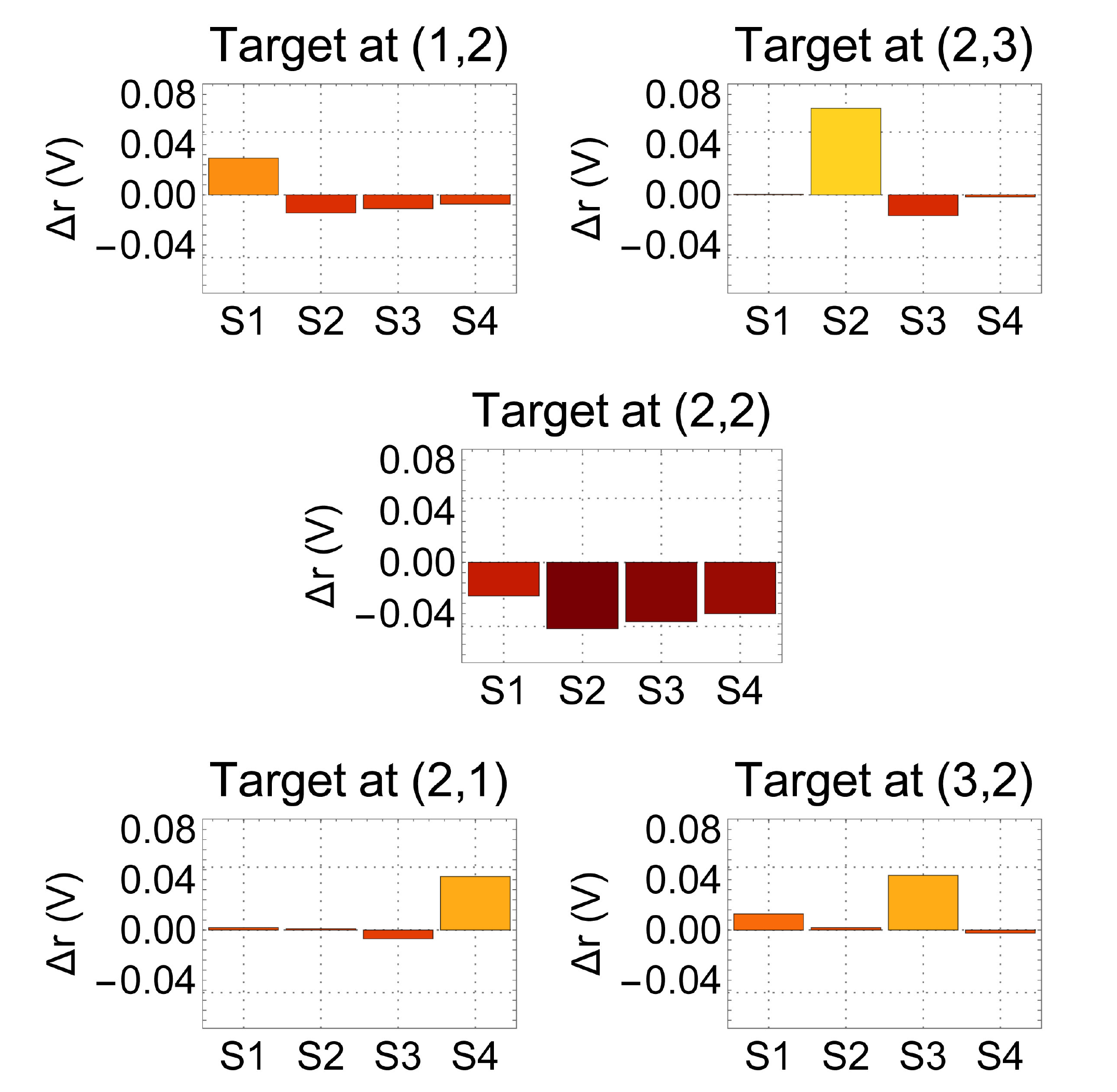}
\caption{Target localization: simultaneously recorded $\Delta$r, when an Al plate (105$\times$110$\times$10~mm$^{3}$), in air 90~mm above the sensors' plane, is placed in different positions. Operation frequency: 20~kHz.}
\label{fig:sensors-r}
\end{figure}

This is demonstrated in Fig.~\ref{fig:sensors-r}. $\Delta$r is \emph{simultaneously} measured by the four Sn when the Al plate is placed in different positions, 90~mm above the sensors' plane. The LabVIEW control displays the measured values in real time and compares them to the corresponding background. An alarm showing the target's position on the grid is automatically triggered upon detection (see also Visualization 1 \cite{video}).

Clear signatures of the target's presence and position are confirmed, with negligible cross-talk among Sn. We attribute the small variations observed in some cases to random fluctuations \textcolor{black}{($\leq$10$\%$ in the response of the same sensor in the same conditions, over several days, in the case of the smallest tested target sized 44$\times$50$\times$13~mm$^{3}$)}. However, these do not hamper the detection and localization of the target: we obtained 100$\%$ correct \emph{automatic} detections and localizations with the Al plate.

The phase variation $\Delta \phi_{n}=\phi_{target}-\phi_{bg}$ is also recorded. A decrease is observed when the object is above a sensor, leading to unambiguous detection and correct localization. \textcolor{black}{After detailed analysis, we found $\Delta \phi$ to be a less robust parameter than $\Delta$r. We attribute this to the larger intrinsic variability of the phase data. Therefore, $\Delta \phi$ is not taken into further consideration in this work}.

\begin{figure}[htbp]
\centering
\includegraphics[width=\linewidth]{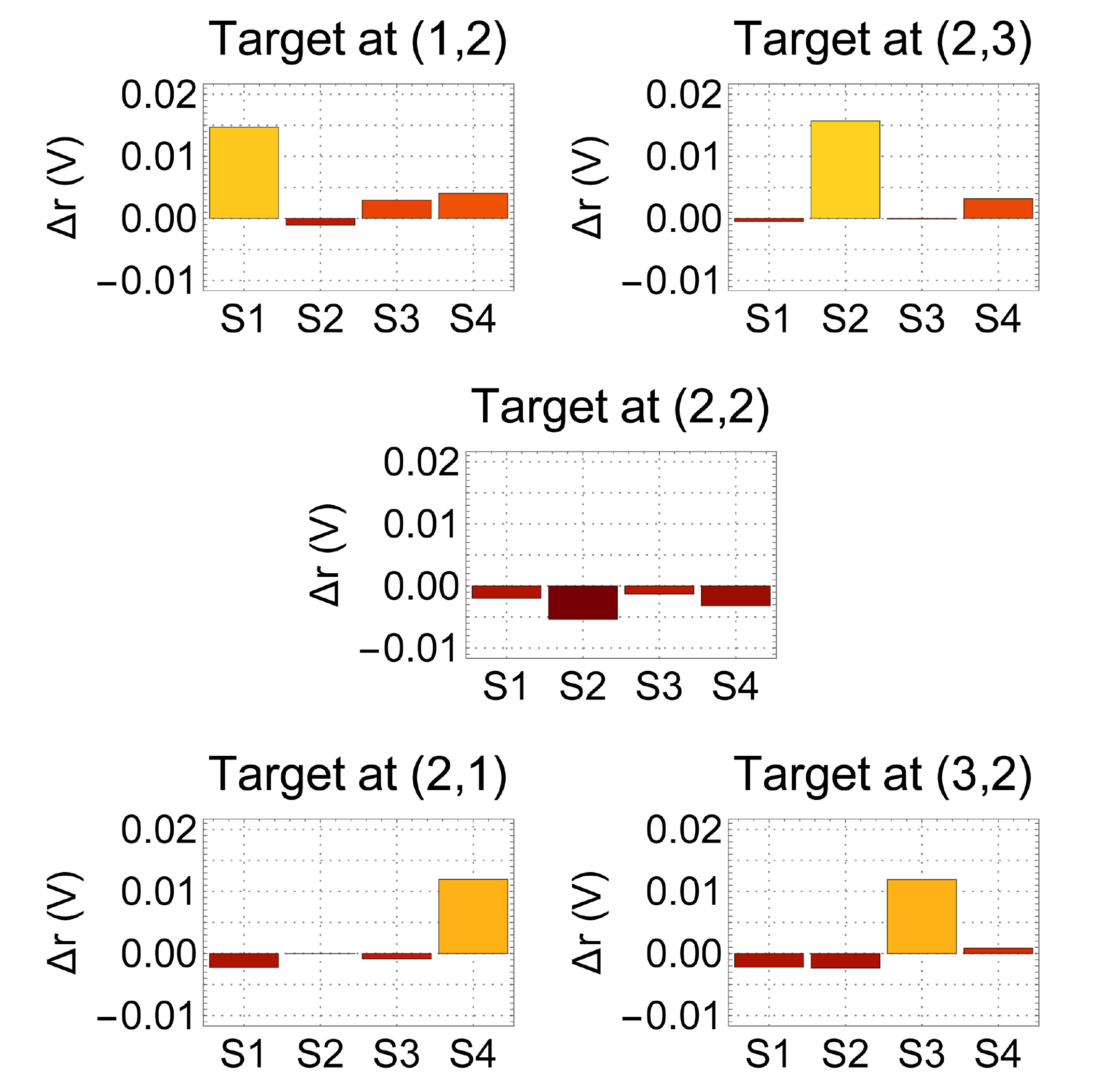}
\caption{Target localization: simultaneously recorded $\Delta$r, when an Al target (44$\times$50$\times$13~mm$^{3}$) in air, 90~mm above the sensors' plane, is placed in different positions. Operation frequency: 20~kHz.}
\label{fig:sensors-r-holes}
\end{figure}

Figure~\ref{fig:sensors-r-holes} \textcolor{black}{shows} the results of a similar experiment, conducted with an Al block of 44$\times$50$\times$13~mm$^{3}$ in air, 90~mm above the sensors' plane. A five-fold decrease of the signal is observed. This is due to the smaller size of the target: the 5$\times$ decrease in  $\Delta$r is consistent with the ratio of the two targets' areas, 5.25 \cite{footnote}.

Nevertheless, clear detection and effective localization are demonstrated. Overall, a correct localization rate of 95.2$\%$ over 21 sets of tests was obtained at 20~kHz.

\section{Underwater Detection and Localization}\label{sec:sea}
\subsection{Saline Water Test-Bed}\label{subsec:underwater}
To test the array with underwater detection and localization, we used a test-bed mimicking the worst-case scenario of sea water. A 25~mm thick Perspex platform above the main coil supports a Nylon water tank (355$\times$215$\times$265~mm$^{3}$). The target is immersed in the water and held at different depths. The sensors' plane and the water level are separated by 90~mm containing air and two layers of plastic. This mimics the realistic scenario of a sensing platform above water on a supporting structure and an underwater target.

To reproduce the average electric conductivity of sea water, we used a 0.0231 NaCl solution with de-ionized water. De-ionized water allows detailed control of the solution conductivity. A NaCl/water solution with a salinity of 22.1 matches the sea conditions at 22$^{\circ}$C: $\varepsilon_{sea}=\varepsilon_{0}\varepsilon_{r,sea}=8.854\cdot10^{-12} \cdot 80$~F/m, $\mu=\mu_{0}\mu_{r,sea}=1.26\cdot 10^{-6}\cdot 1$~H/m, $\sigma$=3.3~S/m \cite{seaconductivity, sea}.

In the band chosen for this work (10 to 20~kHz), skin depth in sea water varies between 2.8~m and 1.9~m. Penetration of km can be achieved by further reducing the RF frequency (Eq.~\ref{eqn:delta}).

\subsection{Underwater Experimental Results}
In this section, we demonstrate the detection and localization of underwater non-magnetic targets.

\begin{figure}[htbp]
\centering
\includegraphics[width=\linewidth]{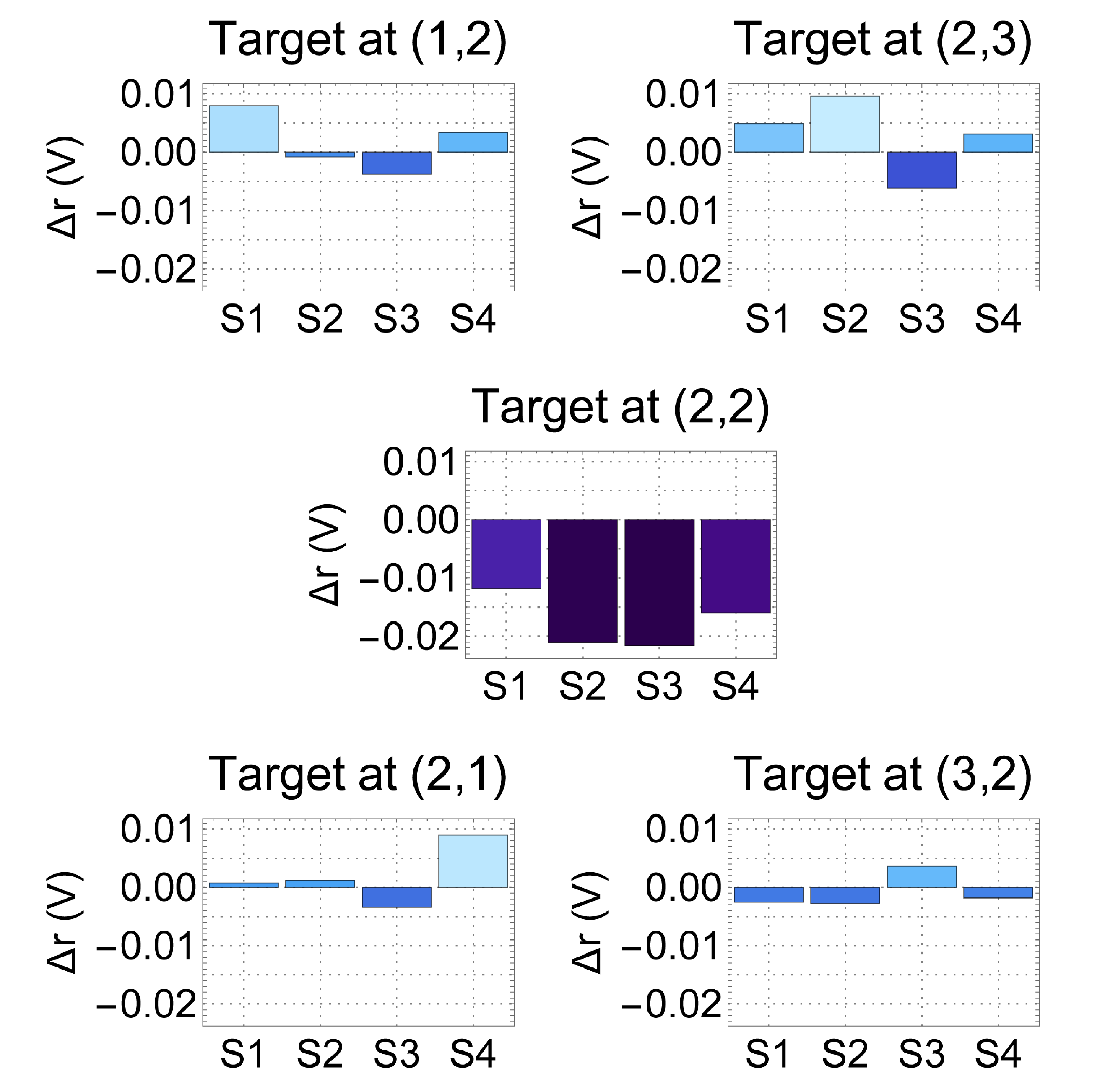}
\caption{Underwater target detection and localization: simultaneously recorded $\Delta$r, when an Al plate (105$\times$110$\times$10~mm$^{3}$) is placed in different positions, at 30~mm underwater (120~mm from the array plane). Operation frequency: 10~kHz.}
\label{fig:water-block-30mm}
\end{figure}

In Fig.~\ref{fig:water-block-30mm}, the Al plate is placed \textcolor{black}{30~mm} under water, 120~mm above the array's plane.

The absolute levels of the amplitude signals decrease. In particular, we measured $\Delta$r $\sim$8 times smaller than Fig.~\ref{fig:sensors-r}. This value reduces to a factor $\sim$3 in the case of the central $(2,2)$ position. Nevertheless, the array measures consistent variations in presence of the target, and unambiguous detection and localization is achieved.

Overall, excellent success rates for automatic localization are obtained: over 22 separate tests at different depths, the success rate was 91$\%$. This number increases to 95$\%$ for depths smaller than 50~mm ($\leq$140~mm above the array plane). At deeper locations, the success rate progressively decreases: \textcolor{black}{at 65~mm deep}, automatic localization is successful in 70$\%$ of cases. \textcolor{black}{No correlations were found between the position and the failure rate.}

\begin{figure}[htbp]
\centering
\includegraphics[width=\linewidth]{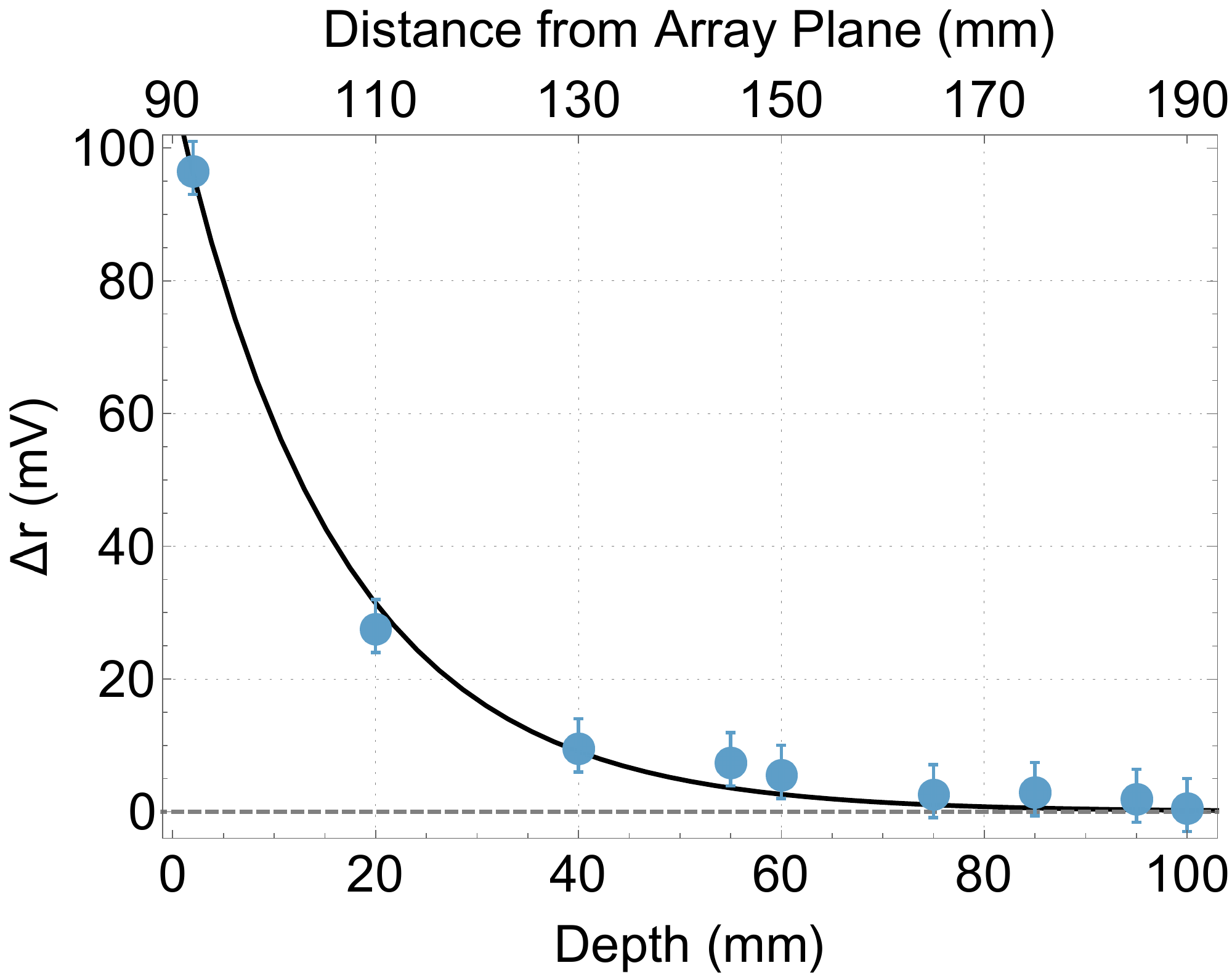}
\caption{Underwater target detection: $\Delta$r of S2 as a function of the depth of the target, a thin Al plate (105$\times$73$\times$3~mm$^{3}$), detected at 10~kHz, placed in position (2,3), above the sensor. \textcolor{black}{The dashed horizontal line marks the ``zero-level'' compatible with no possible detection, corresponding to 100 mm underwater (190 mm distance from the sensors' plane)}.}
\label{fig:penetration}
\end{figure}

Underwater depth and distance from the RF coil play a relevant role, as demonstrated by Fig.~\ref{fig:penetration}. In the graph, S2 $\Delta$r is plotted versus the depth underwater and the distance from the sensors' plane of a thin Al plate (105$\times$73$\times$3~mm$^{3}$) in position $(2,3)$.

Water attenuates both the primary and the secondary fields, producing a noticeable decrease of the MIT signal. Such variation is estimated of the order of $\sim$10$\%$. Therefore, water attenuation alone cannot account for the observed decrease in $\Delta$r. These results could be improved by optimizing the design of the RF source, or by using better approaches for automatic localization, such as machine learning \cite{ml}. \textcolor{black}{Increasing the number of sensors will also improve the localization and tracking performance.} \textcolor{black}{We recall that detection and localization in shallow water - where acoustic methods are overwhelmed by echoes and optical methods are hampered by opacity and turbulence of water - is of primary importance for a number of fields, including surveillance, industrial monitoring, and surveying.}

Finally, by using the LabVIEW automatic real-time detection interface, demonstrations of: i) Detection and localization of multiple targets; ii) Suppression of background structures or targets; and iii) Live tracking of moving targets were demonstrated, as shown in Visualization 1 \cite{video}. These features are essential pre-requisites for the proposed applications, in particular surveillance. The active nature of the MIT approach makes conventional passive countermeasures such as magnetic de-gaussing \cite{degauss}, or acoustic cloaking \cite{cloaking} ineffective. We also note that, given the broad tunability of AMs, the primary field source frequency can be easily tuned for achieving different penetrations, as well as avoiding noisy bands, or concealing the active probing.

\section{Conclusions}
We have demonstrated detection and localization of conductive, non-magnetic targets by using a 2$\times$2, planar array of radio-frequency atomic magnetometers operating in magnetic induction tomography configuration. The sensors in the array are operated continuously and simultaneously with negligible cross-talk. The active nature of the magnetic induction detection, combined with the sensitivity and tunability of RF AMs, make this approach suitable for operation in unshielded environments, \textcolor{black}{and} achieving long penetration ranges, to detect fixed and moving targets.

A proof-of-concept demonstration in \textcolor{black}{saline water} proved the feasibility of an MIT-based array of AMs for underwater detection and tracking, also in shallow water where other conventional approaches have limited effectiveness.

Our results lay the grounds for potential applications from archeological surveys to civil engineering. \textcolor{black}{We found no evidence to suggest fundamental limitations in scaling up the array or increasing the monitored area.} Furthermore, integrating the active tracking demonstrated here with passive detection of the AC electromagnetic signatures of motors \cite{ao} could provide a novel class of early surveillance platforms, suitable for land, air, and water applications.

\vskip 5pt
UK Quantum Technology Hub in Sensing and Metrology, Engineering and Physical Sciences Research Council (EPSRC) (EP/M013294/1). Engineering and Physical Sciences Research Council (EPSRC) (EP/L015242/1).

\vskip 5pt
Cameron Deans \textcolor{black}{acknowledges} support by the EPSRC.

%


\begin{thebibliography}{32}
\bibitem [{foo()}]{footnote}%
  \BibitemOpen
  \href@noop {} {}\bibinfo {note} {At 20~kHz, $\delta_{Al}$=0.58~mm, which is
  much smaller that the thickness of both targets. Consequently, the relevant
  parameter for eddy current in this case is only the surface enclosing the
  magnetic flux changes. This corresponds, for each of the two targets, to
  their areas.}\BibitemShut {Stop}%
\end{thebibliography}
\end{document}